\documentclass[11pt,a4paper]{article}

\usepackage[margin=2 cm,twoside]{geometry}
\usepackage[english]{babel}
\usepackage[utf8]{inputenc}
\usepackage[T1]{fontenc}
\usepackage{siunitx} 
\usepackage{url} 
\usepackage{fancyhdr} 
\usepackage{lastpage} 
\usepackage[notlof,notlot,nottoc]{tocbibind}

\usepackage[font=footnotesize,labelfont=bf]{caption}

\usepackage{float}
\usepackage{amsmath} 
\usepackage{amssymb} 
\usepackage{graphicx} 
\usepackage{color} 
\usepackage{subcaption}
\usepackage{physics} 
\usepackage{cite}
\usepackage{url} 
\usepackage[hidelinks]{hyperref}

\begin{document}
\vspace*{1cm}
\begin{center}
    {\Large 
    \textbf{Nonlocal Electro-Optic Metasurfaces for Free-Space Light Modulation}}
    \\
    {\large \vspace{0.5cm}
    Christopher Damgaard-Carstensen\textsuperscript{*}, and Sergey I. Bozhevolnyi\textsuperscript{*}} \\
    \textit{Centre for Nano Optics, University of Southern Denmark, Campusvej 55, \\ DK-5230 Odense M, Denmark} \\[0.5cm]
    \textsuperscript{*}To whom correspondence should be addressed: \\
    cdc@mci.sdu.dk, seib@mci.sdu.dk
\end{center}

\vspace{1cm}

\noindent \textbf{Abstract}:
Dynamic optical metasurfaces with ultrafast temporal response, i.e., spatiotemporal optical metasurfaces, provide attractive solutions and open fascinating perspectives for modern highly integrated optics and photonics. In this work, electro-optically controlled optical metasurfaces operating in reflection and utilizing resonant waveguide mode excitation are demonstrated from the viewpoint of free-space propagating light modulation. The modulation of reflected light power with superior characteristics in comparison with prior research is achieved by identifying a suitable low-loss waveguide mode and exploiting its resonant excitation. The electro-optic Pockels effect in a 300-nm-thick lithium niobate (LN) film sandwiched between a continuous thick gold film and an array of gold nanostripes, serving also as control electrodes, is exploited to realize fast and efficient light modulation. The fabricated compact (active area <1000 µm\textsuperscript{2}) modulators operate in the wavelength range of 850–950 nm, featuring a maximum intensity modulation depth of $\sim\!$ 42 \% at the driving voltage of $\pm$10 V within the bandwidth of 13.5 MHz (with the potential bandwidth of 6.5 GHz). The introduced nonlocal electro-optic metasurface configuration opens new avenues towards the realization of ultrafast, efficient and robust free-space light modulators based on an LN flat optics approach.

\section{Introduction}
Optical metasurfaces, comprising planar arrays of nanoscale resonant subwavelength elements, constitute one of the most promising and rapidly developing fields in optics and photonics, primarily due to their ability of exercising complete control over the transmitted and reflected fields \cite{Ding2017,Hsiao2017,Chen2016,Yu2011}. Large flexibility offered by optical metasurfaces has enabled the design and realization of metasurfaces featuring diverse optical functionalities, including beam steering \cite{Pors2013_BS,DamgaardCarstensen2020_BS,Li2015_BS}, focusing with planar lenses \cite{Yi2017_PL,Ding2019_PL,Boroviks2017_PL,Engelberg2020_PL}, and generation of optical holograms \cite{Wen2015_OH,Chen2013_OH,Huang2015_OH}. One limitation to the available applications for metasurfaces is the static nature of most metasurface configurations, meaning that their optical responses are determined in the design stage and set in the process of fabrication. In recent years, however, increasing efforts have been made to realize dynamic (tunable) optical metasurfaces \cite{Sinatkas2021,Shalaginov2020,Che2020}. These developments will unlock the metasurface applications in current and emerging technologies, e.g. in light detection and ranging (LIDAR) \cite{Schwarz2010}, spatial light modulators (SLMs) \cite{Smolyaninov2019,BeneaChelmus2021,Park2020_SLM}, computational imaging and sensing \cite{Jung2018}, and virtual and augmented reality systems \cite{Shaltout2019}. When using metasurfaces, the achievable interaction length is severely limited, due to the fundamentally thin nature of metasurfaces, thus making the realization of efficient dynamic optical metasurfaces more challenging. Significant modulation can be realized, despite the low interaction length, by exploiting materials allowing for large refractive index changes, e.g. phase change materials \cite{Shirmanesh2020,Park2020,Wang2021,Zhang2021,Abdollahramezani2022}, and materials with large thermo-optic effect \cite{Sharma2020,Rahmani2017}, or structural reconfigurations \cite{She2018,Li2016}, however, these are all inherently slow. A different approach, which can offer outstanding tuning of the optical path length by actuating static metasurfaces, are MEMS configurations, although their switching speed is still limited because of the necessity of mechanical movements \cite{Meng2021,Arbabi2018,Meng2022}. 

Several ferroelectric media, like lithium niobate (LN) \cite{Thomaschewski2020}, electro-optic polymers \cite{Zhang2018,BeneaChelmus2022}, aluminum nitride \cite{Smolyaninov2019}, and lead zirconate titanate (PZT) \cite{Alexander2018}, exhibit the linear electro-optic (Pockels) effect, which offers inherently fast electrical control over refractive index changes. LN is an appealing platform for dynamic optical components \cite{Zhang2021_LNreview}, due to high electro-optic coefficients ($r_{33}$ = 31.45 pm/V for the extraordinary polarization and $r_{13}$ = 10.12 pm/V for the ordinary one \cite{Jazbinek2002}), a wide transparency range (0.35 - 4.5 µm), excellent chemical and mechanical stability, and a large Curie temperature of $\sim\!$ 1200 \textsuperscript{o}C \cite{Weis1985}. The main drawback of this platform is inherently very small refractive index variations that, along with the previously mentioned limitation in the available interaction length, results in low modulation depths \cite{Gao2021,Weigand2021}. The effective interaction length can however be significantly increased by using resonance configurations (i.e., by multiple usage of the same interaction length), thereby increasing the modulation depth, although its increase inevitably occurs at the expense of a decreased operating wavelength range \cite{DamgaardCarstensen2021,Weiss2022,DamgaardCarstensen2022,Weigand2021}. 

Here, we build on our previous work on dynamic flat optics components based on a thin film LN platform, in which we exploited the polarization-independent Fabry-Perot resonant configuration \cite{DamgaardCarstensen2021} and the resonant excitation of TM-polarized (the electrical field is in the propagation plane) waveguide mode in a LN thin film sandwiched between an array of parallel subwavelength-spaced gold nanostripes and a optically thick gold back-reflector \cite{DamgaardCarstensen2022}. With the latter (most recently explored) configuration used for free-space propagating light modulation, we have demonstrated the modulation depth of 20 \% for $\pm$10 V of driving voltage \cite{DamgaardCarstensen2022}. We note that hybridization of the resonant excitation of TM-polarized waveguide mode and the Fabry-Perot resonance allows for the possibility of polarization independent operation \cite{DamgaardCarstensen2022}. To achieve more efficient modulation (without targeting polarization independent operation), one way would be to increase the refractive index modulation, e.g., by increasing the modulation voltage applied or the Pockels coefficient used. An alternative way is to utilize a narrower resonance, which is equivalent to increasing the effective interaction length. In this case, it will require a smaller shift of the resonance to achieve a large intensity difference. In the investigation of our previous modulator, we noticed a much sharper resonance within the same parameter range, namely the resonance associated with the excitation of a low-loss TE-polarized (the electrical field is perpendicular to the propagation plane) waveguide mode. The grating excitation of guided modes in the LN thin film is equivalent to a strongly nonlocal response, since it requires synchronization of scattering in the metasurface plane, and the resonant excitations can propagate in-plane as guided modes \cite{Zhang2022,Zhang2020,Kwon2018,Song2021}. Since the waveguide mode propagation constant depends on the LN refractive index, the resonance wavelength can be tuned via the electro-optic effect by applying an electric field. In this work, we make use of this sharp resonance and realize a significantly improved electro-optic free-space modulator in otherwise the same LN-based configuration. By means of numerical calculations and extensive experiments, we design, fabricate, and characterize free-space intensity modulators operating in the wavelength range of 850-950 nm. The modulation depth of $\sim\!$ 40 \% at the driving voltage of $\pm$10 V is demonstrated within the bandwidth of 13.5 MHz, thus offering characteristics that are favorable compared to those achieved previously \cite{DamgaardCarstensen2022,Weiss2022}. The introduced nonlocal electro-optic metasurface configuration opens new avenues towards the realization of ultrafast, efficient and robust free-space light modulators based on an LN flat optics approach.

\section{Design and simulations}

\begin{figure}[htbp]
	\centering
	\includegraphics{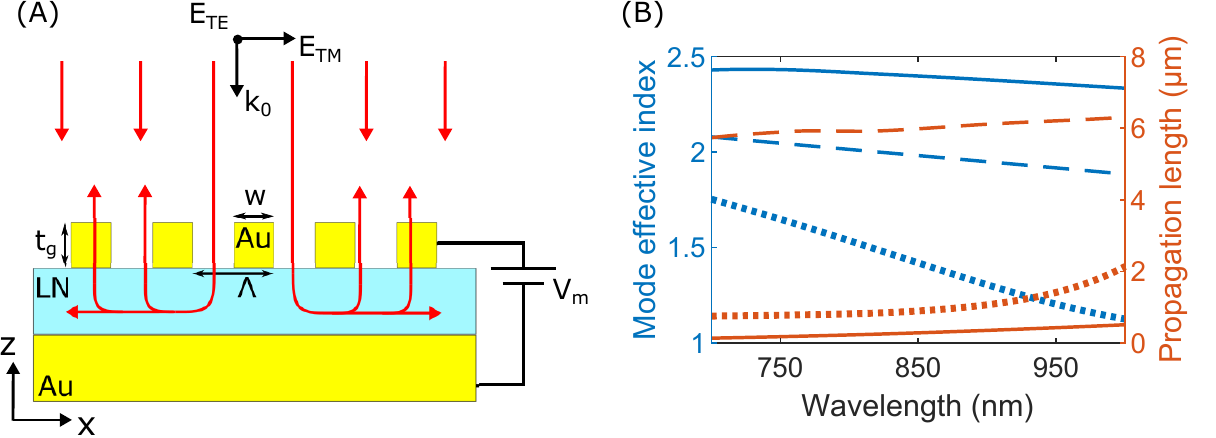}
	\caption{Grating excitation of nonlocal modes. (A) Cross-section schematic of the device configuration. The modulator consists of subwavelength spaced gold nanostripes patterned on a LN thin film adhered to an optically thick gold back-reflector by a thin chromium layer. The structure is supported by 3 µm of silicon dioxide and a bulk LN substrate. A modulation voltage is applied between the gold back-reflector and nanostripes. (B) Dispersion relations for the three fundamental modes supported by the LN film in the absence of periodic gold stripes on top, namely a fundamental plasmonic TM mode (solid), a fundamental TE mode (dashed), and a second TM mode (dotted).}
	\label{fig1}
\end{figure}

The considered material configuration exploited in this work is essentially the same as those used in our previous work \cite{DamgaardCarstensen2021,DamgaardCarstensen2022} and consists of a custom material stack from NANOLN: LN substrate, 3 µm of SiO\textsubscript{2}, 30 nm of chromium, 300 nm gold, 10 nm of chromium, and a 300 nm $z$-cut LN thin film. Chromium layers are introduced for adhesive purposes only, introducing additional absorption loss \cite{DamgaardCarstensen2022} that would otherwise be desirable to avoid. In contrast to the prior work however, we exploit the resonant excitation of a low-loss TE-polarized waveguide mode supported by the LN thin film. To excite this waveguide mode, we employ the previously used fork configuration, which consists of subwavelength-spaced, parallel, semi-infinite gold nanostripes connected at one end, thus forming a grating [Fig. \ref{fig1}(A)]. 

The gold nanostripes and back-reflector function also as integrated metal electrodes for tuning, thus allowing for a very compact device. An electric field is formed in the sandwiched LN thin film, when a voltage is applied between the bottom and top electrodes. This stimulates the Pockels effect and prompts a shift in refractive index, which, for an electric field along the optical axis ($z$-axis), is given by the Pockels first-order derivation \cite{Thomaschewski2022}: 
\begin{equation} \label{eq:Pockels}
	|\Delta n| \simeq \frac{1}{2} n^3 r_{hk} \frac{V}{t_{LN}} 
\end{equation}
where $n$ is the refractive index with no voltage applied, $r_{hk}$ is the relevant Pockels coefficient, $t_{LN}$ is the LN film thickness, and $V$ is the modulation voltage. Given that we excite a guided TE mode, the electric field component is in the metasurface plane, and thus we should use the Pockels coefficient $r_{13} = 10.12$ pm/V \cite{Jazbinek2002} and the ordinary refractive index, $n_o$.

\begin{figure}[htbp]
	\centering
	\includegraphics{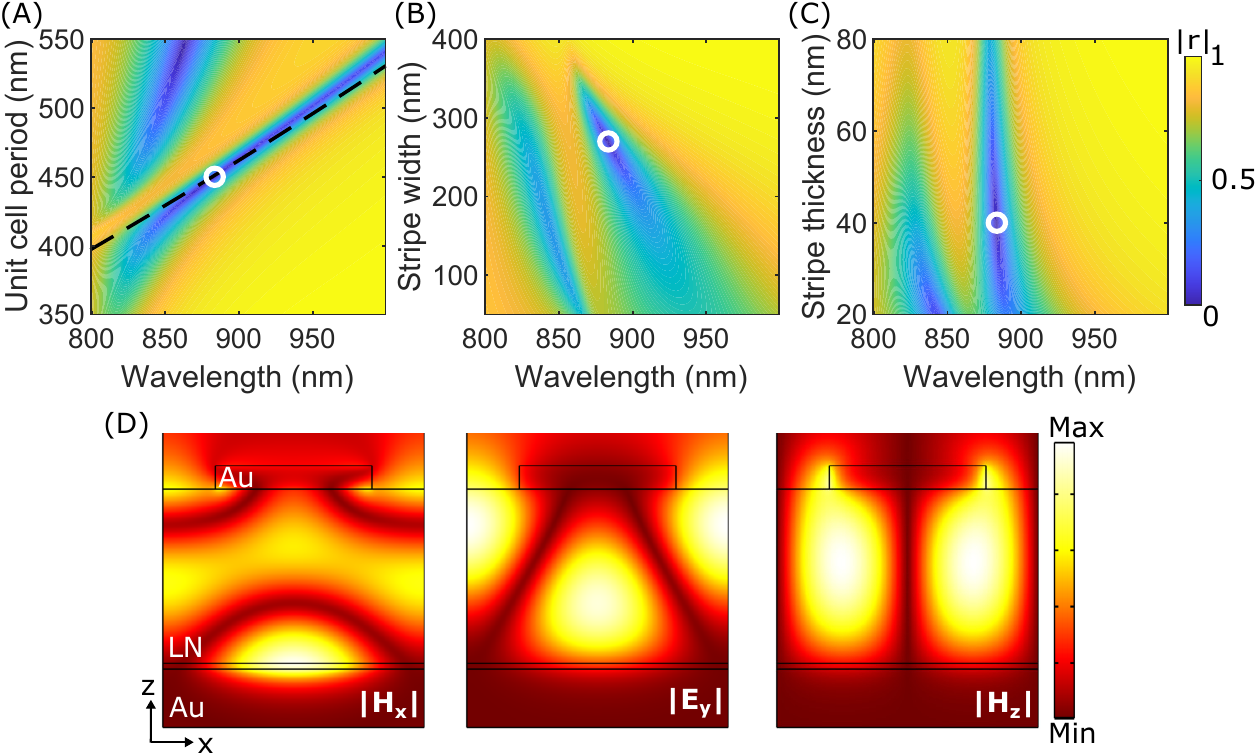}
	\caption{Design and optimization of the nonlocal metasurface. (A-C) Colormaps of simulated reflection amplitude for varying wavelength and (A) unit cell period, $\Lambda$, (B) stripe width, $w$, and (C) stripe thickness, $t_g$. White circles represent the selected design parameters, which are: $t_g$ = 40 nm, $\Lambda$ = 450 nm, $w$ = 270 nm, and $t_{LN}$ = 300 nm. (A) The black dashed line is calculated from the dispersion of the fundamental TE mode in the LN thin film. (D) Field profiles at resonance for a TE polarized normally incident wave.}
	\label{fig2}
\end{figure}

To achieve strong modulation, one should look for a sharp and deep resonance that can be advantageously exploited for the electro-optic modulation. A deep resonance can be achieved at critical coupling, where scattering and absorption losses are equal, and the result is complete radiation absorption \cite{Wu2011}. To simultaneously achieve a sharp resonance, one should strive to reduce the losses, meanwhile keeping the absorption and scattering losses equal to maintain critical coupling. The absorption loss is controlled by the propagating waveguide mode, and thus to reduce this loss channel we look for a low-loss mode. The scattering loss is controlled by the stripe parameters, i.e. width and thickness, which we have full control over. As a result, we first look for a low loss mode, and afterwards optimize the stripe parameters to achieve critical coupling. Since our configuration has a metal substrate any TM modes will naturally be more lossy, and for that reason we look for TE-polarized modes.
Initially, we characterize the first three modes supported by the LN film in the absence of periodic gold stripes on top [Fig. \ref{fig1}(B)]. The configuration supports a fundamental plasmonic TM mode (solid lines), a fundamental TE mode (dashed lines), and a second TM mode (dotted lines), with the TE waveguide mode clearly displaying the largest propagation length (smallest loss). Taking into account the periodic gold stripes on the top, we calculate a propagation length of 3 µm at critical coupling, corresponding to half the calculated propagation length of Fig. \ref{fig1}(B) and equivalent to more than six grating periods, confirming the nonlocal response. 

The design procedure can be explained by following the reasoning developed in our latest work \cite{DamgaardCarstensen2022} and reflected in the panels shown in Fig. \ref{fig2}(A-C). Among different resonant branches seen in these panels that correspond to different physical mechanisms involved in the resonant absorption, we have chosen the narrowest resonant branch, which was identified as corresponding to the resonant excitation of the TE-polarized waveguide mode. The dispersion of the fundamental TE mode supported by the LN film in the absence of periodic gold stripes on its top was used to calculate the stripe array periodicity required for the excitation of this mode under normal light incidence for different wavelengths. The corresponding dependence is shown as the black dashed line in Fig. \ref{fig2}(A), demonstrating that indeed this resonant branch corresponds to the TE-polarized waveguide mode excitation. Slight deviation observed for wavelengths larger than 900 nm is related to the presence of the periodic gold stripes atop the LN thin film, that modifies the TE mode propagation constant. For wavelengths shorter than 850 nm, a very strong perturbation caused by hybridization of this waveguide resonance with the Fabry-Perot resonance is observed [Fig. \ref{fig2}(A)]. The Fabry-Perot resonance should not (strongly) depend on the array period, but when these two resonant dispersion branches intersect, one observes a typical anti-crossing behavior caused by the strong coupling, leading to the repulsion of the resonant dispersion curves \cite{Trm2014}. 

To optimize the nonlocal resonance, we iteratively sweep the available design parameters one at a time. The LN thickness is set in the custom stack, thus leaving three available design parameters, namely, the stripe width, stripe thickness, and unit cell period. The unit cell period and stripe thickness are initially set to 450 nm and 40 nm to ensure excitation at a desired wavelength and a homogeneous gold film. Subsequently, the stripe width is swept to determine a deep and narrow resonance. Each parameter influences the resonance differently, with the variation in stripe width and thickness being less influential, and the variation in unit cell period having the biggest influence [Fig. \ref{fig2}(A-C)]. White circles represent the optimized design parameters of: $w$ = 270 nm, $t_g$ = 40 nm, and $\Lambda$ = 450 nm. Field distributions of the final design at resonance show a field confined in the LN thin film for maximum tunability and the interference pattern of an in-plane travelling mode, thus supporting that we are working around the resonant excitation of a waveguide mode [Fig. \ref{fig2}(D)]. 

\begin{figure}[htbp]
	\centering
	\includegraphics{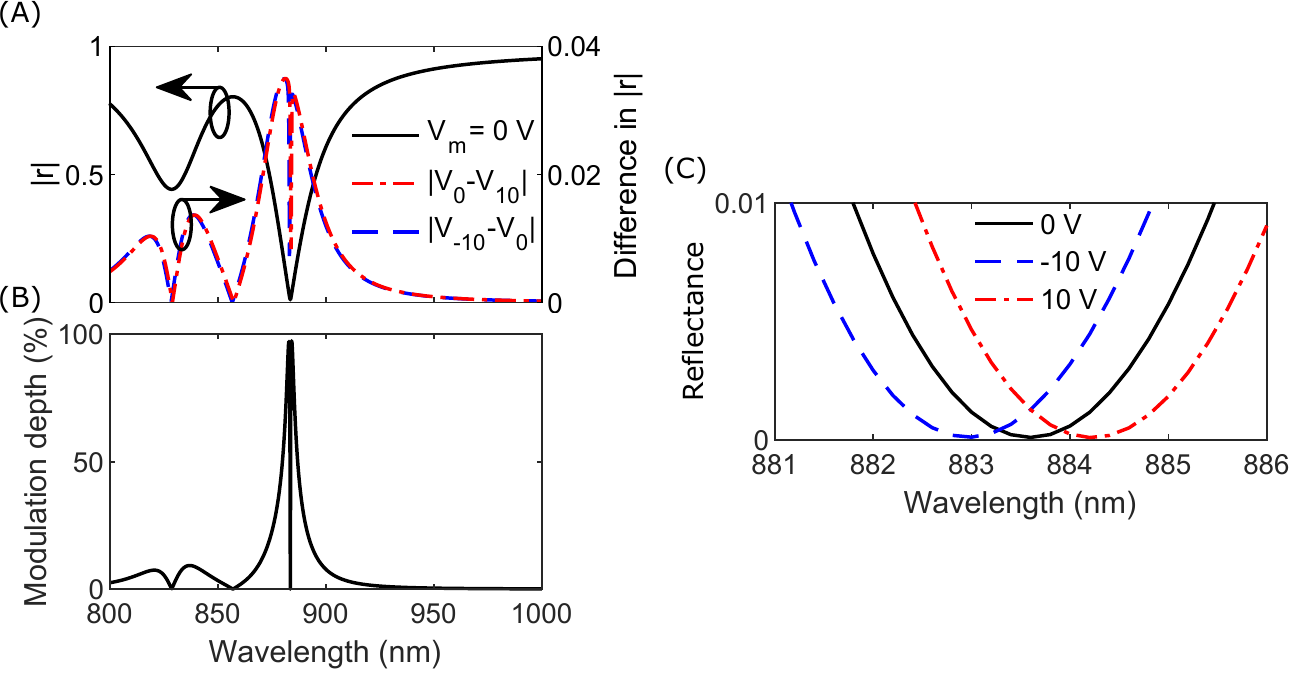}
	\caption{Calculated performance of the nonlocal metasurface. (A) Calculated reflection amplitude for a passive device (left axis) and difference in amplitude when applying a modulation voltage of $\pm$ 10 V (right axis). (B) Corresponding calculated modulation depth from the results of (a). (C) Zoom-in plot displaying the resonance shift due to an applied modulation voltage.}
	\label{fig3}
\end{figure}

A primary characteristic of intensity modulators, is the modulation depth, given by $1-(|r_{min}(\lambda)|^2$ $/|r_{max}(\lambda)|^2)$, where $r_{min}(\lambda)$ and $r_{max}(\lambda)$ are the minimum and maximum reflection amplitudes, respectively \cite{Yao2014}. By applying a modulation voltage of $\pm$10 V, we observe a calculated difference in reflection amplitude around the resonance [Fig. \ref{fig3}(A)]. From the equation for calculating modulation depth and the complete radiation absorption at critical coupling, we expect and observe a calculated modulation depth, which extends up to 100 \% at resonance [Fig. \ref{fig3}(B)]. The desired deep and sharp resonance is present at a wavelength of 884 nm, and a zoom-in view of the reflectance at resonance visualizes the wavelength shift of 1.2 nm [Fig. \ref{fig3}(C)]. This wavelength shift results in a tuning sensitivity of $\Delta\lambda/\Delta V = 0.06$ nm/V. 
A larger modulation depth can be achieved by increasing the spectral shift of the resonance or by utilizing a sharper resonance, described by the full width half maximum (FWHM). To quantify the modulation capability, we calculate a figure of merit (FOM) based on these qualities for our current and latest works \cite{DamgaardCarstensen2022}, $\mathrm{FOM}=\Delta\lambda / \mathrm{FWHM}$. Despite that our current work exhibits a smaller wavelength shift (1.2 nm vs. 2.0 nm for our latest work \cite{DamgaardCarstensen2022}), we achieve a FOM almost 50 \% larger (0.046 for our current work and 0.033 for our latest work), because the resonance is sharper.

\section{Fabrication and characterization}

\begin{figure}[htbp]
	\centering
	\includegraphics{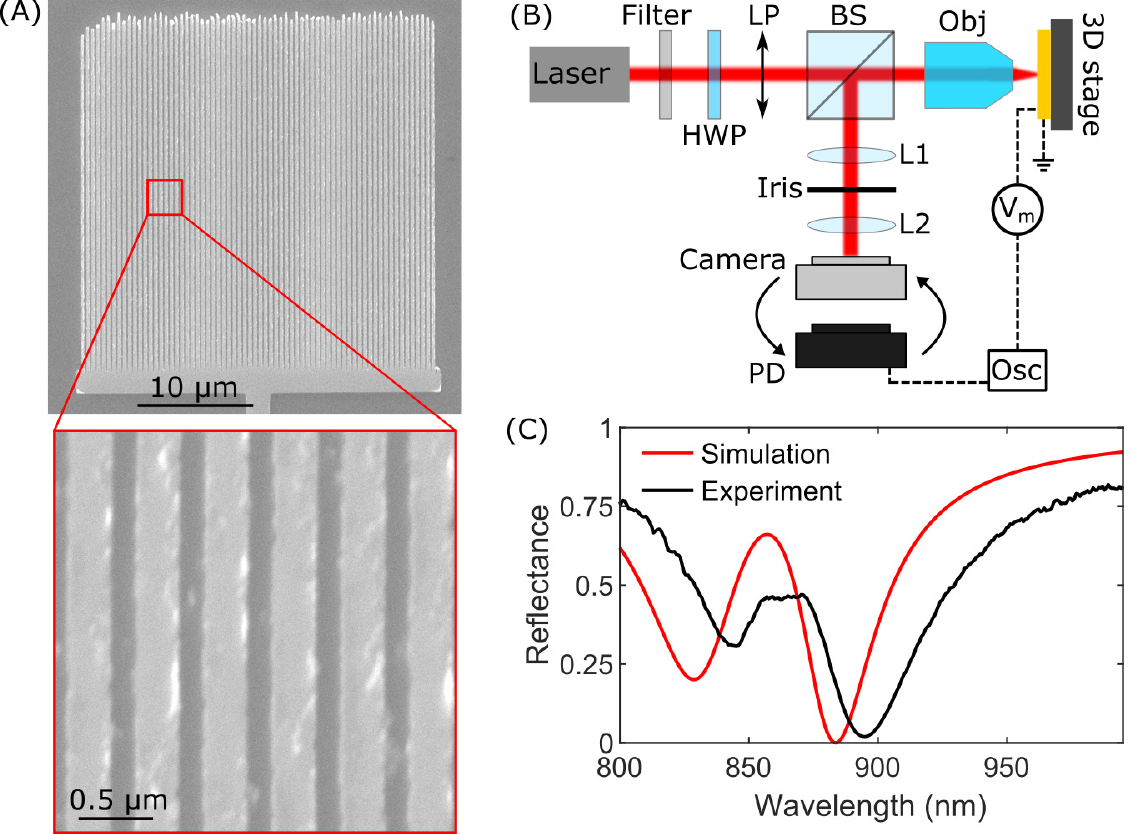}
	\caption{Imaging and characterization of a passive nonlocal metasurface. (A) SEM image of an entire modulator as well as a zoom-in view of six nanostripes. (B) Setup used for electro-optical characterization. Red lines indicate the optical path, and black dashed lines indicate the electrical wires. HWP: Half-wave plate, LP: Linear polarizer, BS: Beam splitter, Obj: Objective, L1(2): Lens 1(2), PD: Photodetector, Osc: Oscilloscope, V\textsubscript{m}: Modulation voltage source. (C) Measured reflection spectrum of the modulator (black line) and simulated spectrum (red line), normalized to reflection from plane gold. Errors are in the size range of line widths.}
	\label{fig4}
\end{figure}

After having conducted thorough numerical calculations, the designed nonlocal metasurface is fabricated using the conventional technique of electron-beam lithography and lift-off (see Methods). Scanning electron microscopy images after fabrication show evenly-spaced straight nanostripes with only minor defects at the edge of the grating [Fig. \ref{fig4}(A)]. During fabrication, the modulator is interfaced by a 2 µm wide gold wire to macroscopic electrodes for electrical connection. For optical characterization of the passive device, a laser beam from a supercontinuum source is linearly polarized and slightly focused by a 20X objective to form a beam waist that is slightly smaller than the modulator area [Fig. \ref{fig4}(B)]. The reflected light is collected by the same objective and separated from the incident light by a beam splitter, whereafter it is imaged to a plane with an iris, which is used for spatial filtering, before it is imaged again to a camera or detected by a spectrometer. The measured and calculated spectra show similar wavelength dependence, however the measured spectrum is slightly red-shifted [Fig. \ref{fig4}(C)]. The reason for this may be slight fabrication inaccuracies or the titanium adhesion layer. The requirement to utilize a resonant configuration to achieve a larger effective interaction length, and thus a larger modulation depth, results in an insertion loss of -15.2 dB at maximum modulation depth for our nonlocal metasurface, with insertion loss defined as $10 \log_{10} (|r_{max}(\lambda)|^2) $ \cite{Yao2014}. 

Lateral miniaturization is directly related to the achievable resolution and device footprint for implementation in ultrathin and ultrafast SLMs. To quantify this, we calculate the minimal achievable pixel size by extracting the decay length of the optical mode at critical coupling \cite{Weiss2022,DamgaardCarstensen2022}. For our fabricated device the loaded Q-factor is approximately 30, which gives an unloaded Q-factor of $Q_U=2Q_L\approx 60$ at resonance, where the two loss channels of absorption and coupling loss are equal due to critical coupling. From the unloaded Q-factor we estimate the loss per unit length: 
\begin{equation}
	\alpha = \frac{4\pi}{Q_U\lambda_{eff}} \approx 0.47 \mathrm{\mu m^{-1}}
\end{equation}
where $\lambda_{eff}$ is the effective wavelength, and we assume an effective refractive index of 2. The inverse of the loss per unit length gives the propagation length of $L\approx 1/0.47 \mathrm{\mu m^{-1}} = 2.1 \mathrm{\mu m}$, thus corresponding well with our calculated propagation length of 3 µm. We note that for implementation in an SLM configuration the pixel size can be considerably reduced without risking cross-talk due to guided modes. 

\begin{figure}[htbp]
	\centering
	\includegraphics{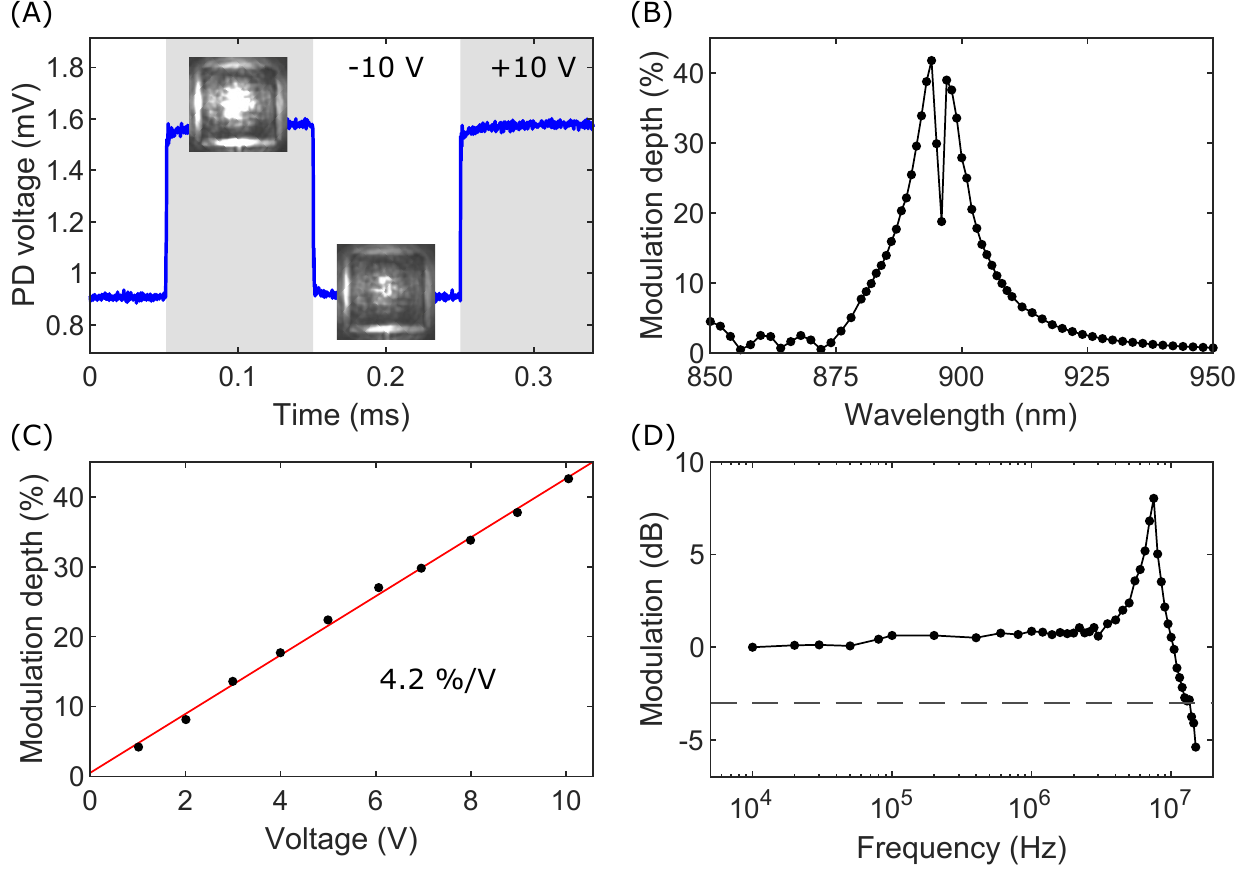}
	\caption{Characterization of an active nonlocal metasurface. (A) Measured intensity with a photodetector (PD),  visualizing the modulation when cycling a square voltage signal between +10 V and -10 V indicated by grey and white backgrounds, respectively. Insets: Optical images of the device for a modulation voltage of +10 V (upper left image) and -10 V (lower right image), clearly displaying the modulation of reflected intensity. (B) Experimentally measured modulation depth for varying wavelengths at a modulation voltage of $\pm$10 V and signal frequency of 5 kHz. (C) Experimentally measured modulation depth for increasing modulation voltage (black dots) at a signal frequency of 5 kHz. Indicated voltages represent amplitudes of the applied AC modulation signal. A linear fit to the data points (red line) shows a modulation increase of 4.2 \%/V. (D) Experimentally measured normalized modulation for increasing signal frequency. The dashed line represents -3 dB. All data points are normalized to the lowest applied frequency of 10 kHz. For all plots, error bars are in the order of data point sizes.}
	\label{fig5}
\end{figure}

Efficient modulator operation relies on the efficient excitation of a waveguide mode with a periodic array of nanostripes and requires therefore the incident laser beam to have a perfectly flat phase profile, i.e., representing a well-defined spatial mode, which is difficult to obtain from supercontinuum laser sources not based on high finesse resonators, contrary to, for example, tunable Ti:Sapphire lasers, which we employ for characterization of the modulator operation. Driving voltages are supplied by a function generator, and the reflected light is detected by a photodetector in otherwise the same optical setup [Fig. \ref{fig4}(B)]. Initially, we visualize the modulation by measuring the intensity change when cycling a square voltage signal between +10 V and -10 V [Fig. \ref{fig5}(A) and Visualization 1]. Insets show optical images of the modulator at $\pm$10 V. For a similar driving voltage, we measure the modulation depth in the wavelength range of 850-950 nm, achieving a maximum modulation depth of 42 \% at a wavelength of 895 nm [Fig. \ref{fig5}(B)]. The shape of the measured wavelength dependence corresponds well with calculated results, but the maximum modulation is reduced by a factor of 2.5 [Fig. \ref{fig3}(B)]. This reduction is believed to be primarily due to fabrication inaccuracies and the titanium adhesion layer \cite{DamgaardCarstensen2022}. Further, we note that a small amount of background radiation or an imperfect resonance will considerably reduce the measured modulation depth, because the very high calculated modulation depth is largely due to the very deep resonance. The expected linear relation between modulation depth and voltage, because of the linear Pockels effect, is verified by a linear fit to the measured data points, which gives an increase in modulation depth of 4.2 \%/V [Fig. \ref{fig5}(C)]. The electrical bandwidth is another primary characteristic of a modulator, and in the frequency range of 10 kHz to 15 MHz we plot normalized modulation [Fig. \ref{fig5}(D)] and observe a -3 dB cutoff frequency of 13.5 MHz. We note increases in the modulation for larger frequencies, which we attribute to piezoelectric resonances in LN and changes in permittivity and electro-optic coefficients following the transition to the clamped crystal response \cite{Thomaschewski2020,Jazbinek2002,Takeda2012}. Based on a measured capacitance of 0.30 nF and assuming a 50 $\Omega$ resistive load, an estimate of the cutoff frequency is calculated to be 10.6 MHz. The difference to the measured cutoff frequency is believed to be due to the previously mentioned reasons for the increase in modulation at larger frequencies. The capacitance of the electrical feedline on the chip is the main limitation on the switching speed of our device, with the electro-optic Pockels effect easily supporting much higher bandwidths. If the bottom electrode is limited to an area only below the patterned modulator, our device demonstrates a calculated capacitance of 0.49 pF and cutoff frequency of 6.5 GHz, which is easily supported by the Pockels effect. 

\section{Conclusion}
To summarize, we have presented and experimentally characterized an optical intensity modulator based on a LN thin film sandwiched between an optically thick bottom and nanostructured top gold films, serving also as control electrodes. The fabricated compact (active area <1000 µm\textsuperscript{2}) device demonstrates effective electro-optic tuning of the resonant excitation of a TE-polarized waveguide mode for compact modulation of the reflectance of light. The modulation depth reaches $\sim\!$ 40 \% for the modulation voltage of $\pm$10 V within the electrical bandwidth of 13.5 MHz (with the potential bandwidth of $\sim\!$ 6.5 GHz) for near-infrared radiation (850-950 nm). Several free-space metasurface-based intensity modulators have been published in recent years, utilizing various active materials (Table \ref{tab:Comparison}). The configuration presented here is attractive because of using the material (LN) having an inherently fast electro-optic response, being superior to modulators based on phase change materials or MEMS configurations, and excellent environmental stability, contrary to electro-optic polymer-based modulators with a glass temperature below 100 \textsuperscript{o}C. Overall, we believe the introduced nonlocal electro-optic metasurface configuration opens new avenues towards the realization of ultrafast, efficient, and robust free-space light modulators based on an LN flat optics approach. 

\begin{table}[htbp]
	\centering
	\scriptsize
	\setlength{\tabcolsep}{1.7pt}
	\caption{Comparison of recently published works on electrically tunable free-space modulators with our work, comparing several key characteristics. Modulation efficiency is defined as the product of modulation depth and reflectance/transmittance. Abbreviations are: IR: Infrared, NIR: Near-infrared, PCM: Phase-change material, GST/GSST: Germanium antimony telluride alloys, EOP: Electro-optic polymer, LN: Lithium niobate, AlOx: Aluminum oxide.}
	\label{tab:Comparison}
	\begin{tabular}{ccccccccc}
		\textbf{Active}   & \textbf{Active} & \textbf{Modulation} & \textbf{Reflectance/}       & \textbf{Modulation} & \textbf{Electrical}      & \textbf{Driving}  & \textbf{Wavelength} & \textbf{Minimum} \\[-3pt]
		
		\textbf{material} & \textbf{thickness} & \textbf{depth} & \textbf{Transmittance} & \textbf{efficiency} & \textbf{bandwidth} & \textbf{voltage} & \textbf{range}      & \textbf{pixel size} \\[-3pt]
		
		& \textbf{(nm)} & & & & \textbf{(MHz)} & \textbf{(V)} &  & \textbf{(µm)} \\ \hline
		
		AlOx/Graphene \cite{Yao2014} & 300 & 95 \% & 50 \% & 47.5 \% & 40 & 0-80 & Mid-IR & - \\
		
		GaAs/AlGaAs \cite{Pirotta2021} & 368 & 30 \% & 50 \% & 15 \% & 750 & 0-6 & Mid-IR & - \\
		
		PCM (GSST) \cite{Zhang2021} & 250 & 75 \% & 45 \% & 34 \% & - & 0-23 & Telecom & - \\
		
		PCM (GST) \cite{Wang2021} & 25 & 77 \% & 30 \% & 23 \% & 0.01 & - & Visible-NIR & - \\
		
		PCM (GST) \cite{Abdollahramezani2022} & 40 & 88 \% & 85 \% & 75 \% & 0.0035 & 0-4 & Telecom & - \\
		
		EOP (JRD1:PMMA) \cite{BeneaChelmus2021} & 690 & 37 \% & 85 \% & 31 \% & 50 & $\pm 80$ & Telecom & - \\ 
		
		EOP (JRD1:PMMA) \cite{BeneaChelmus2022} & 600 & 67 \% & 90 \% & 60 \% & 3000 & $\pm 100$ & Telecom & - \\
		
		LN \cite{Weigand2021} & 500 & 0.012 \% & 45 \% & 0.0054 \%  & 2.5 & $\pm 5$ & Visible-NIR & - \\ 
		
		LN \cite{Weiss2022} & 480 & 40 \% & - & - & 0.778 & $\pm 25$ & Telecom & 8.3 \\ 
		
		LN \cite{DamgaardCarstensen2022} & 320 & 20 \% & 4 \% & 0.8 \% & 8.5 & $\pm 10$ & NIR & 1.1 \\ 
		
		\textbf{LN (This work)} & \textbf{300} & \textbf{42 \%} & \textbf{3 \%} & \textbf{1.3 \%} & \textbf{13.5} & $\mathbf{\pm 10}$ & \textbf{NIR} & \textbf{2.1} \\ 
		
	\end{tabular}
\end{table}

\section{Methods}

\subsection{Modeling}
The commercially available finite element software \textit{COMSOL Multiphysics}, ver. 5.6 is used for simulations of the device performance. All simulations are performed for 2D models, because the grating design is constant in the $y$-direction. Interpolated experimental values are used for the permittivity of gold \cite{Johnson1972}, chromium \cite{Johnson1974}, and LN \cite{Zelmon1997}. Due to the significant thickness of the chromium adhesion layer (10 nm), and our inability to alter it, we include it in the optical simulations, even though adhesion layers are typically not modelled. The design is periodic, and thus a single period with one nanostripe is modeled and periodic boundary conditions are added to the side walls of the cell. Ports are applied to the top and bottom boundaries to minimize reflection and handle wave excitation and measure a complex reflection coefficient. The top port is positioned one wavelength from the top of the nanostripe, and the incident light is a plane wave travelling downward, normal to the sample. A simulation is performed in two steps: First, the electric field distribution from an applied DC voltage is determined in an electrostatic simulation. Second, the change of refractive index is calculated from the electrid field distribution and the Pockels coefficients of Jazbin{\v{s}}ek et. al. \cite{Jazbinek2002} (considering only the largest diagonal terms, i.e., $\Delta n_i = -\frac{1}{2} n_i^3 r_{iiz} E_z$, with $r_{xxz}=r_{yyz} =$ 10.12 pm/V and $r_{zzz} =$ 31.45 pm/V), after which the optical simulation is conducted with the updated refractive index. For calculation of the mode-effective index, we employ the mode solver of \textit{COMSOL Multiphysics}. 

\subsection{Fabrication}
The device is fabricated using a combination of nanostenciling and electron beam lithography. 5 nm of titanium and 100 nm of gold is deposited by thermal evaporation (Cryofox Tornado 400) through a shadow mask to form macroscopic electrodes. A $\sim\!$ 200 nm layer of PMMA 950K A4 resist is spin-coated, and the modulator is manually aligned to the macroscopic electrode and exposed using electron beam lithography at 30 kV (JEOL JSM-6490LV equipped with an Elphy Quantum lithography system). The resist is developed and the modulator is formed by evaporation of 3 nm of titanium and 40 nm of gold followed by liftoff in acetone. The square nanostripe array consists of 70 periods, and thus it is 31.5 µm $\times$ 31.5 µm. The useful volume of the device, taking into account the active thickness of 0.3 µm, is 31.5 µm $\times$ 31.5 µm $\times$ 0.3 µm $\simeq$ 298 µm\textsuperscript{3}.

\subsection{Characterization}
Connection to the upper macroscopic electrode is obtained by mounting the sample on a homemade sample holder, and connection to the bottom electrode by applying a conductive paste to the edge of the sample. The sample holder is mounted on a 3D stage. 
For characterization of the passive device, a collimated supercontinuum laser beam (NKT Photonics SuperK Extreme) is used in combination with a spectrometer to get the full spectral response, and for characterization of the active device, a collimated low-power continuous-wave laser beam from a tunable laser (Spectra-Physics 3900 S Ti:Sapphire) is used. The reason is that efficient modulator operation relies on the efficient excitation of a waveguide mode with a periodic array of nanostripes and requires therefore the incident laser beam to have a perfectly flat phase profile, i.e., representing a well-defined spatial mode, which is difficult to obtain from supercontinuum laser sources not based on high finesse resonators (contrary to, for example, tunable Ti:Sapphire lasers). 
The incident light is polarized by a rotatable linear polarizer and focused by a 20X objective (Mitutoyo M Plan Apo 20X 0.42) to form a spot size smaller than the modulator area. Reflected light is collected by the same objective, separated from the incident light by a beam splitter, and imaged to a plane where it is spatially filtered by an iris, after which it is imaged again to a camera (Thorlabs DMM1545M-GL) or detected by a photodetector (Thorlabs PDA20CS-EC or PDA10A2) or spectrometer (Ocean Optics Optem QEPRO). AC modulation signals are supplied by a function generator (Toellner TOE 7401 or Hewlett Packard 8648C). Modulation performance is measured by an oscilloscope (Keysight InfiniiVision DSOX2024A) connected to the function generator and photodetector.

\subsection*{Funding}
  The authors acknowledge financial support from Villum Fonden (Award in Technical and Natural Sciences 2019). DOI: 10.13039/100008398

\subsubsection*{Supplementary Material}
See Visualization 1 for supporting content.

\end{document}